\documentclass[12pt]{article}
\usepackage[dvips]{graphicx}
\usepackage{rotating}

\usepackage{epsfig,amsmath,amssymb,verbatim,mathrsfs}
\numberwithin{equation}{section}
\def\beq{\begin{eqnarray}}
\def\eeq{\end{eqnarray}}
\def\bea{\begin{eqnarray}}
\def\eea{\end{eqnarray}}

\newcommand{\gsim}{\lower.7ex\hbox{$\;\stackrel{\textstyle>}{\sim}\;$}}
\newcommand{\lsim}{\lower.7ex\hbox{$\;\stackrel{\textstyle<}{\sim}\;$}}

\def\stilde{\widetilde}

\newcommand{\newc}{\newcommand}
\newc{\Nc}{N_{c}}
\newc{\CG}{C_G}
\newc{\gp}{g'}
\newc{\stopi}{\stilde t_i}
\newc{\sboti}{\stilde b_i}
\newc{\staui}{\stilde \tau_i}
\newc{\stopj}{\stilde t_j}
\newc{\sbotj}{\stilde b_j}
\newc{\stauj}{\stilde \tau_j}
\newc{\stopI}{\stilde t_1}
\newc{\stopII}{\stilde t_2}
\newc{\sbotI}{\stilde b_1}
\newc{\sbotII}{\stilde b_2}
\newc{\stauI}{\stilde \tau_1}
\newc{\stauII}{\stilde \tau_2}
\newc{\sstop}{s_{t}}
\newc{\cstop}{c_{t}}
\newc{\ssbot}{s_{b}}
\newc{\csbot}{c_{b}}
\newc{\sstau}{s_{\tau}}
\newc{\cstau}{c_{\tau}}
\newc{\Sstop}{s_{2t}}
\newc{\Cstop}{c_{2t}}
\newc{\Ssbot}{s_{2b}}
\newc{\Csbot}{c_{2b}}
\newc{\Sstau}{s_{2\tau}}
\newc{\Cstau}{c_{2\tau}}
\newc{\salpha}{s_\alpha}
\newc{\calpha}{c_\alpha}
\newc{\Calpha}{c_{2\alpha}}
\newc{\Salpha}{s_{2\alpha}}

\newc{\sbetapm}{s_{\beta_\pm}}
\newc{\cbetapm}{c_{\beta_\pm}}
\newc{\Sbetapm}{s_{2 \beta_\pm}}
\newc{\Cbetapm}{c_{2 \beta_\pm}}
\newc{\sbetaO}{s_{\beta_0}}
\newc{\cbetaO}{c_{\beta_0}}
\newc{\SbetaO}{s_{2 \beta_0}}
\newc{\CbetaO}{c_{2 \beta_0}}
\newc{\vu}{v_u}
\newc{\vd}{v_d}
\newc{\seL}{\stilde e_L}
\newc{\smuL}{\stilde \mu_L}
\newc{\seR}{\stilde e_R}
\newc{\smuR}{\stilde \mu_R}
\newc{\suL}{\stilde u_L}
\newc{\sdL}{\stilde d_L}
\newc{\suR}{\stilde u_R}
\newc{\sdR}{\stilde d_R}
\newc{\scL}{\stilde c_L}
\newc{\ssL}{\stilde s_L}
\newc{\scR}{\stilde c_R}
\newc{\ssR}{\stilde s_R}
\newc{\snue}{\stilde \nu_e}
\newc{\snumu}{\stilde \nu_\mu}
\newc{\snutau}{\stilde \nu_\tau}
\newc{\Gpm}{G^\pm}
\newc{\Hpm}{H^\pm}
\newc{\FFbS}{\overline{FF}S}
\newc{\FFbV}{\overline{FF}V}
\newc{\FSS}{F_{SS}}
\newc{\FSSS}{F_{SSS}}
\newc{\FFFS}{F_{FFS}}
\newc{\FFFbS}{F_{\overline{FF}S}}
\newc{\FSSV}{F_{SSV}}
\newc{\FVS}{F_{VS}}
\newc{\FVVS}{F_{VVS}}
\newc{\FFFV}{F_{FFV}}
\newc{\FFFbV}{F_{\overline{FF}V}}
\newc{\Fgauge}{F_{\rm gauge}}
\newc{\DRbarprime}{$\overline{\rm DR}'$ }
\newc{\DRbar}{$\overline{\rm DR}$ }
\newc{\MSbar}{$\overline{\rm MS}$ }
\newc{\Yu}{{\bf Y}_u}
\newc{\Yd}{{\bf Y}_d}
\newc{\Ye}{{\bf Y}_e}
\newc{\Au}{{\bf a}_u}
\newc{\Ad}{{\bf a}_d}
\newc{\Ae}{{\bf a}_e}
\newc{\bm}{{\bf m}}
\newc{\zhol}{Z^{\rm hol}}

\newcommand{\Ebar}{\overline{E}}
\newcommand{\dnde}{\frac{dn}{dE_e}}
\newcommand{\ecut}{E_{\rm cut}}

\begin{document}

\setlength{\baselineskip}{0.2in}

%#!latex

%\begin{comment}

%@twocolumnfalse\endcsname
%%
%%
\begin{titlepage}
\noindent
\begin{flushright}
MCTP-09-12\\
%HUTP-08-xx\\
\end{flushright}
\vspace{1cm}

\begin{center}
  \begin{Large}
    \begin{bf}
Pulsars as a Source of the WMAP Haze \\
     \end{bf}
  \end{Large}
\end{center}
\vspace{0.2cm}

\begin{center}

\begin{large}
Manoj Kaplinghat$^{a}$, Daniel J. Phalen$^{b}$,
and Kathryn M. Zurek$^{b,c}$\\
\end{large}
\vspace{0.3cm}
  \begin{it}
$^a$ Center for Cosmology, Department of Physics and Astronomy\\
University of California, Irvine, CA, USA\\
\vspace{0.5cm}
${^b}$ Department of Physics, University of Michigan\\
Ann Arbor, Michigan 48109, USA\\
\vspace{0.5cm}
$^c$ Particle Astrophysics Center, Fermi National Accelerator Laboratory\\
Batavia, Illinois 60510, USA

\end{it}

\end{center}

\center{\today}

%\maketitle
\begin{abstract}

The WMAP haze is an excess in the 22 to 93 GHz frequency bands of WMAP 
extending about 10 degrees from the galactic center. We show that
synchrotron emission from electron-positron pairs injected into the
interstellar medium by the galactic population of pulsars with
energies in the 1 to 100  GeV range can explain the frequency spectrum
of the WMAP haze and the drop in the average haze power with
latitude. The same spectrum of high energy electron-positron pairs
from pulsars, which gives rise to the haze, may also generate the
observed excesses in AMS, HEAT and PAMELA. We discuss the spatial
morphology of the pulsar synchrotron signal and its deviation from
spherical symmetry, which may provide an avenue to determine the pulsar
contribution to the haze.
\end{abstract}

\vspace{1cm}

\end{titlepage}
%\pacs{PACS numbers: }
%]

%\setcounter{footnote}{1}
\setcounter{page}{2}
%\setcounter{figure}{0}
%\setcounter{table}{0}

%\tableofcontents

\vfill\eject

%\end{comment}

%%%%%%%%%%%%%%%%%%%%%%%%%%%%%%%%%%%%%%%%%%%%%%%%%%%%%%%%%%%%%%%%%%%%%%

\newpage

\section{Introduction}

The measurements of the temperature anisotropies in the Cosmic
Microwave Background (CMB) by the Wilkinson Microwave Anisotropy Probe
(WMAP) have made possible an impressively precise extraction of
cosmological parameters.  WMAP has also been able to measure
Interstellar Medium (ISM) emission of low energy photons from dust and 
ionized gas in the inner regions of the galaxy.  However, an excess of
synchrotron radiation in the WMAP bands between 22 and 93 GHz has
been observed, after a subtraction of the free-free, dust and standard
synchrotron emission \cite{Finkbeiner2,Dobler:2007wv}.  This
excess has been dubbed the WMAP ``haze.''   

One possible explanation of the excess is in terms of annihilating
dark matter \cite{Finkbeiner3}.  The charged byproducts of the
dark matter annihilations radiate synchrotron photons in the
galactic magnetic field.  A neutralino from supersymmetric theories
annihilating to $W^+W^-$ gives a good fit to the radial distribution
of the spectrum for a dark matter halo profile scaling with a radial
dependence which is slightly steeper than NFW \cite{FinkbeinerHooper}
and a magnetic field in the few $\mu$G range.  The cross-section of
annihilating dark matter needed to produce the haze is consistent with
what one would predict from the thermal freeze-out of the WIMP, namely
$\sigma v \simeq 3 \times 10^{-26} \mbox{ cm}^3/\mbox{s}$. 
We urge caution, however, because recent
work~\cite{Cumberbatch:2009ji} has claimed that the significance of   
the WMAP haze may depend on the assumptions about the spatial
variation of the synchrotron spectral index.
 
This signal may be even more interesting in light of the recent
observations of an excess of high energy cosmic ray positrons and
electrons.  The excess was originally observed by the HEAT \cite{HEAT}
and AMS \cite{AMS} experiments, and was confirmed more recently by the
PAMELA \cite{PAMELA} and PPB-BETs \cite{PPB-BETs}
experiments.  The Fermi Large Area Telescope also observes excess cosmic ray electrons and positrons over the predicted background \cite{Fermi}, though their excess is not nearly as large as that observed by ATIC \cite{ATIC} (and seems to indicate that the ATIC excess is instrumental in origin).  The source of these positrons is unknown, however, there
are several possibilities.  Like the WMAP haze, they may be explained
by annihilating DM (see {\em e.g.} \cite{Cirelli,HooperWeiner}).  An
explanation of the signal in terms of annihilating dark matter (DM),
however, has multiple obstacles to overcome.  First, it must
annihilate with a cross-section significantly larger than that suggested by the
thermal abundance,  $\sigma v \simeq 3 \times 10^{-24-23} \mbox{
  cm}^3/\mbox{s}$.  Second, the DM candidate prefers leptophilic annihilation to avoid overproducing anti-protons \cite{Cirelli,Italian} and
to produce a steep enough spectrum \cite{HooperWeiner}.  Gamma rays
and radio measurements also generate significant constraints
\cite{Cirelli,Bergstrom}, since the charged SM byproducts of the
annihilation may emit either hard final state radiation or synchrotron
emission in the galactic magnetic field.  In short, if the positron
excesses are to be explained in terms of annihilating Weakly
Interacting Massive Particle, the WIMP must have non-standard
properties.  There are possible exceptions to these conclusions in the
case that we happen to be nearby a dense clump of dark matter
\cite{clump}, or for non-standard propagation models \cite{Michigan}. 

The cosmic ray positron excess may be purely astrophysical, however.
There are several possible astrophysical explanations.  The excess may
result from secondary production and acceleration in the cosmic ray
source itself \cite{BlasiCR}.  Supernova remnants may also produce 
highly energetic positrons to explain the PAMELA signal \cite{Piran}.
It has been noted many times that a single or few pulsars in 
the neighborhood of a kiloparsec from us, or a distribution of pulsars, may
contribute $e^+e^-$ with an energy spectrum that can reproduce the  
steep rise over the $E^{-3}$ background seen in PAMELA
\cite{Atoian:1995ux,Chi:1995id,ChengZhang,Blasi,Yuksel:2008rf,Profumo,Malyshev:2009tw}.

In this paper we explore the possibility that the WMAP haze may also
be generated by $e^+ e^-$ pair production in pulsars.  Pulsars produce
a significant flux of energetic electrons and positrons spread over
the disk of the galaxy, which then emit synchrotron radiation as they
traverse outward from the disk in the magnetic field of the galaxy.
To explain the haze, the expected signal from pulsars must both
reproduce angular dependence of the signal from the galactic center,
as well as the frequency dependence through the WMAP bands from 22 to
93 GHz.  
We show that an energy spectrum which is typical of that
necessary for explaining the PAMELA data with pulsars also naturally
produces the average variation of WMAP haze with latitude given typical galactic magnetic fields and
diffusion parameters. We also discuss reproducing the detailed morphology of the WMAP haze.

In the next section we describe the model for the electron distribution from mature galactic 
pulsars, and in the following section their propagation through the
ISM and the haze calculation. 

\section{Injection spectrum of positrons from pulsars}
 
The mechanism by which pulsars produce electrons and positrons and
details about their energy distribution are not very well
understood. However, the theoretical models reproduce important
characteristics like the observed distribution of spins, 
ages, and photon fluxes from radio to gamma-rays (e.g.,
\cite{Zhang:2004bu}). Here, we wish to 
demonstrate that with a plausible model for pulsar $e^+e^-$
injection spectra (that is consistent with observations), one can
reproduce the WMAP haze. We  begin by reviewing the model we utilize
for the pulsar $e^+ e^-$ injection spectrum.   

We consider only pulsars older than $10^5$ years as potential sources
of the $e^+e^-$ 
pairs that create the haze.  This is based on the expectation that
young pulsars are surrounded by a nebula created by the kinetic energy
released from the supernova explosion (almost $10^{51}$ ergs)  so  
that $e^+e^-$ cannot escape from this nebula until the pulsar is 
sufficiently old. The nebulae have typical sizes in the parsec range, and since a typical pulsar kick at birth is around 
$\sim 500$  km/s, it would take the pulsar thousands of years to
escape the  nebula.   
In addition, the nebulae themselves thin out in tens of thousands of
years.  For the mature pulsars, we will assume that the 
nebulae surrounding pulsars do not play a dominant role in shaping the
energy spectrum and we neglect the  contributions from pairs diffusing
out of younger pulsar nebulae. The younger pulsars are fewer in number
and could contribute significantly to the higher energy end of the
$e^+e^-$ spectrum. However, the bulk of the synchrotron radiation in
the WMAP bands comes from $e^+e^-$ with energies much less than 100
GeV, which justifies our focus on mature pulsars.  

To demonstrate the feasibility of our assertion that pulsars could
explain most of the visible WMAP haze, we follow the Cheng and Zhang
2001 model (CZ01  from here on)~\cite{ChengZhang}, which relies on the
production of highly energetic radiation in the outer magnetosphere
gap of a rapidly  spinning
pulsar~\cite{Cheng:1986qt,Chiang:1994si,zhang97}. In the CZ01 model,
the mean energy of  $e^+ e^-$ injected into the inter-stellar medium
$\Ebar$ is set by its period, $P$, which increases with time. For a  
rotating magnetic dipole (in vacuum) this spin-down is given by
\begin{eqnarray}
P(t) & = & P_0 \left( 1+\Delta \! t/\tau_0 \right)^{1/2}\,,
\label{eq:dipole}\\
\tau_0 & = & 1.35\times 10^4 {\rm yr} 
\left(\frac{P_0}{30 {\rm ms}}\right)^2 
\left(\frac{M}{1.4M_\odot}\right) 
\left(\frac{R}{15 {\rm km}}\right)^4 
\left(\frac{B}{10^{12} {\rm G}}\right)^{-2}\,, 
\label{eq:tau0}
\end{eqnarray}
where $\Delta \! t$ is time since birth of the pulsar, $P_0$ is the
initial period, $M$ is the mass of the pulsar, $R$ its
radius and $B$ is the surface magnetic field. The energy injected into
the pairs all comes from the spin-down and the surface magnetic field
is assumed to be constant. In the CZ01 model,
\begin{equation}\label{eq:Emean}
\Ebar \simeq 44 {\rm GeV} \left(\frac{P}{0.1 {\rm s}}\right)^{-3.6}\left(\frac{B}{10^{12} {\rm G}}\right)^{0.27}\,
\end{equation}
where $P$ is the period of the pulsar at the time of emission and one
of the parameters of the CZ01 model, the fraction of pairs escaping
from the light cylinder, is set to 0.01.  

The CZ01 model converts a fraction of the available spin-down power
for pulsar ages $>10^5~ {\rm yr}$ to pairs.
To set the scale we note that in this model, for $P=0.5~ {\rm s}$ and
$B=10^{12}~ {\rm  G}$, the differential $e^+ e^-$ emission 
rate is $10^{35} /{\rm GeV}/{\rm s}$, with mean energy $1 ~{\rm
  GeV}$. Only the gamma-ray pulsars are assumed to produce $e^+ e^-$ 
pairs in this model and this introduces a $B$ dependent upper-limit on
the period ($P < 0.25 (B/10^{12} {\rm G})^{6/13} {\rm  s}$).

To predict the properties of the pulsar today, we need the
initial period and magnetic field, and also the initial kick that the
nascent neutron star received when the supernova occurred. The CZ01
study includes a Monte Carlo of these and other parameters that result
in present day distributions that are broadly consistent with
observations. 
We note that in the CZ01 model the spatial distribution of the
injected $e^+e^-$ will depend somewhat on the energy range of
interest if pulsars older than about million years contribute
significantly to that energy range. To test this, we repeat the
modeling of CZ01, including a description of motion of 
pulsars in the galactic potential, and compute the final spatial and
energy distribution of positrons. Our result for the energy
distribution of the positrons agrees with CZ01. In addition, we
find that the mean age of the pulsars, weighted by the positron
ejection rate, is of order $10^5$ years for the energy range of
interest. Given the birth velocities, these ages
imply that typical pulsars (contributing significantly to the positron
flux) have only traveled $\sim 100$ parsecs from their birth
place. We thus use the simple approximation that the spatial 
distribution of these pulsars is the same as the initial
pulsar distribution, which in turn tracks that of the young stars in
the stellar disk. Specifically, we adopt~\cite{Pulsardist,sturner96}   
\begin{eqnarray}\label{eq:rho}
\rho(\vec{x}) & = & N^{-1} e^{-r/r_0-|z|/z_0},\quad {\rm where}\\
N & = & 4 \pi z_0 r_0^2 (1-e^{-r_{disk}/r_0}(1+r_{disk}/r_0))\,,\nonumber
\label{eq:pulsardist}
\end{eqnarray} 
where $r_{disk} = 15 \mbox{ kpc}$, $r_0 = 4 \mbox{ kpc}$ and $z_0 =
100 \mbox{ pc}$. 

The distribution of pulsars close to the center is not well
constrained -- consequently the value of
$r_0=4 \mbox{ kpc}$ that we use should be taken as a very rough
estimate. We have chosen this value based on the dynamical models of
the thin disk,  which assign values for $r_0$ in the range of 2 to
4 kpc. We have adopted 4 kpc for our fiducial model and  
later explore the effect on the morphology of making $r_0$ smaller. In
detail, we expect the pulsar distribution to track the galactic star
formation rate in the disk (rather than the thin disk density), but the
observational constraints on the 
star formation in the disk~\cite{Boissier:1999ts} in the inner couple
of kpc are weak.  One may consider using the observed distribution of
pulsars to reconstruct the true pulsar distribution in the galaxy after
correcting for incompleteness. The distances to these pulsars are
estimated through the dispersion measure, and this depends sensitively
on the assumed distribution of electrons. The
uncertainties inherent in this procedure makes it hard to pin down the
distribution of pulsars in the inner galactic
region~\cite{Lorimer:2006qs}. We note that most of the pulsars in the
distribution assumed in Eq.~\ref{eq:rho} are at $r \sim r_0$. This is
consistent with one of the two models advocated to  
explain the observed pulsar distribution in the Parkes multibeam
pulsar survey~\cite{Lorimer:2006qs}.  We do 
not consider a secondary source of pairs from the pulsars in the bulge
given that the bulge is thought to be old and have a low star
formation rate (compared to the disk). 

The CZ01 Monte Carlo predicts that the energy spectrum of the $e^+
e^-$ pairs should be $E^{-1.6}$ above about a GeV up to tens of
GeV. The spectrum drops sharply above $\ecut\sim 100$ GeV. Both these 
features (the slope and cut-off energy) are model dependent and we
discuss the effect of changing these later on. To keep the discussion 
more general, we therefore adopt an energy spectrum (number of $e^+
e^-$ pairs per unit time per unit energy) given by
\begin{equation}
Q(E) = 
\dot{N}_{100}Q_0 f_e 
\left(\frac{E}{\rm GeV}\right)^{-\alpha}e^{-E/\ecut}\,, 
\label{eq:source}
\end{equation}
where our baseline model has $\alpha=1.6$ and $E_{\rm cut} = 100$
GeV, and we allow it to vary later to see how it changes our
results. We have separated out a factor $f_e$, which is the efficiency
of converting the spin-down power of the pulsar into 
$e^+ e^-$ pairs (after an age of $10^5 ~{\rm yr}$), and the factor 
$\dot{N}_{100}$, which is the number of pulsars created every
century. 

The normalization 
$Q_0$ is fixed by the spin down power of all the pulsars,
that is 
$W_0 = N_p \langle I \Omega^2 \dot{P}/P \rangle$ where
$N_p$ is the total number of pulsars and the brackets indicate
averaging over the galactic population. We set $N_p = 0.01
\dot{N}_{100} (T/{\rm yr}-10^5) \simeq 1000 \dot{N}_{100} T/10^5{\rm
  yr}$, where $T$ is the  typical age of the pulsar contributing the
$e^+ e^-$ pairs.  
The normalization condition for $Q_0$ is given by
\begin{eqnarray}
\int \dot{N}_{100} Q_0 E^{-\alpha} e^{-E/\ecut} 
E dE & = & W_0 = N_p\times 6 \times 10^{38} \frac{\rm GeV}{\rm s}
\langle \left(\frac{P}{0.1 {\rm s}}\right)^{-4}
\left(\frac{B}{10^{12}{\rm G}}\right)^2\rangle\,,\\ 
\Rightarrow Q_0 & \approx & 
5 \times 10^{40} {\rm GeV}^{-1}{\rm s}^{-1}
\frac{100^{\alpha-1.6}\Gamma(0.4)}{\Gamma(2-\alpha)}
\left(\frac{100 {\rm GeV}}{\ecut}\right)^{2-\alpha} 
\,, \label{eq:Q0def}
\end{eqnarray}
where $E$ is the electron or positron energy in GeV, and we have
assumed median values for the pulsar mass of $1.4M_\odot$, radius of
15 km, initial period of 20 ms, and surface  magnetic field of
$2\times 10^{12} {\rm G}$. For these values, Eq.~\ref{eq:tau0} shows
that $T\gg \tau_0$ and therefore $\dot{P}/P=1/2T$ for these mature
pulsars.  
This simple estimate for $Q_0$ agrees with the Monte Carlo results of
CZ01, who find $Q(E) = 1.7 \times 10^{39} E^{-1.6}
\exp(-E/80) / {\rm GeV} /{\rm s}$, if we assume $f_e \simeq 0.03$, 
$\dot{N}_{100} = 1$, $\ecut=80 {\rm GeV}$.

It is important to note that $f_e$ is the fraction of spin down {\em
  power} that is injected into pairs {\em after} the assumed maturity 
  age of $10^5~ {\rm yr}$. This efficiency
  $f_e$ is expected to be large since $e^+e^-$ are the lightest 
  electromagnetically coupled fermions. The fraction of total initial
  energy injected into the ISM in pairs is very small,
  $\sim f_e \tau_0/T$.  
This argument shows that if a significant amount of the spin-down
  energy released before the assumed maturity age of $10^5 {\rm yr}$
  were   
to be available in the form of $e^+e^-$ pairs injected into the ISM,
then the required efficiency would be very small. We see this by
  noting that in the approximation that some fraction of all
  of the spin-down energy is injected into the ISM instantaneously in
  the form of $e^+e^-$ with spectrum $E^{-1.6}$ and cut-off 100 GeV,
  we have  $W_0=\dot{N}_{100}(1/2)I\Omega_0^2/100 {\rm yr}$, which
  works out to $Q_0=2\times 10^{41}$ GeV$^{-1}$ s$^{-1}$ if we take
  (conservatively) $(1/2)I\Omega_0^2 = 10^{52} {\rm GeV}$. The
  efficiency required then to get the same normalization as the CZ01
  model is a factor of 4 less. 

The origin of the $E^{-1.6}$ spectrum in the CZ01 model are the
scalings of $\Ebar$ and the spin down power with period $P$. We note
that $dn/dE \propto T /P^4/E^2 \propto 1/P^2/E^2$ where  
we have used the fact that the number of pulsars is proportional to
the age and we have used the approximation $P^2 \propto T$. We include
the $\Ebar$ dependence on $P$ to obtain $dn/dE \propto E^{2/3.6-2}
\propto E^{-1.4}$, somewhat different from the $E^{-1.6}$ scaling
because of the approximations we have made. 

The cut-off in the spectrum around 100 GeV is related to our
assumption that the pulsars have to be approximately 100 kyr or older to
contribute significantly to the haze, and that the mean energy
of pairs injected into the ISM depends on the pulsar period in the
CZ01 model (see Eq.~\ref{eq:Emean}). This estimate of the cut-off is
uncertain both because of our blanket assumption that pulsars younger 
than 100 kyr do not contribute pairs, and also because in framing the
arguments above we have assumed all pulsars are born with spin
period of 30 ms. We certainly expect scatter about both these
parameters. Including such scatter will change the details of the
cut-off significantly but not the main result of the paper. In
addition, a small change in the strong dependence of the mean energy
on the period would affect the cut-off significantly. This steep
dependence arises from processes that accelerate the pairs into the ISM
and these processes are not well-understood. We note that the energy
spectrum at lower energies ($\sim 10 {\rm GeV}$) is much less
sensitive to the above uncertainties. 

At even lower energies, $E < 1$~GeV, pair emission
luminosity is influenced by the modeling of a gamma-ray pulsar. This
requirement, discussed above, imposes a magnetic field dependent upper
limit on the period, which translates to a cut-off in the luminosity
at low energies. These model assumptions therefore introduce
uncertainties in the GeV range pair emission luminosity, which must be
kept in mind when comparing to experiments like HEAT.

The estimates in this section assuming a vacuum dipole rotator
model for the mature pulsars have provided us with the basic features
of the positron injection spectrum. We find that the spatial 
distribution should track that of the young stars in the disk, with an
energy spectrum that is less steep than $E^{-2}$ -- specifically
$E^{-1.6}$ for the model of CZ01 -- and a total normalization that
requires about $10\%$ of the spin-down power of ${\cal O}(10^5{\rm
  yr})$ pulsars to be injected into positrons.  

\section{Pulsars as a Source for the Haze \label{sec:pulsarmodel}}

The positrons pumped into the ISM will lose energy and diffuse
outwards, and as they do so, they will produce the synchrotron
background. To model this we use the standard diffusion equation that
describes the propagation and energy loss for a charged particle: 
\begin{equation}
\frac{\partial}{\partial t} \dnde(\vec{x},t,E)= K(E)\nabla^2\dnde + \frac{\partial}{\partial E} B(\vec{x},E)\dnde + Q(\vec{x},t,E)\,,
\end{equation}
where $Q(\vec{x},t,E)$
is the source function, i.e., rate of
production of positrons per unit energy with energy $E$ at time $t$
and location $\vec{x}$.  It is a sum over all the pulsars in the
galaxy.  
We have assumed that the diffusion coefficient
$K(E)$ is spatially constant, as there is no evidence in the cosmic
ray or diffuse galactic gamma-ray data to the contrary. In addition,
we note that very little is known about the diffusion constant at 
the center of the galaxy. We will discuss the effect of changing the
diffusion constant on the morphology of the haze later.

The synchrotron emission from the positrons along the line of sight
receives contributions from a large number of pulsars. Since pulsars are
being created on time scales of 100 years in the galaxy and this is
much shorter than the diffusion time scale and the assumed $10^5$ yr
time lag, we expect a steady state calculation to work 
well. In the limit that the diffusion equation can be solved in a
steady state, the source function reduces to 
\begin{equation}
Q(\vec{x},E)= \rho(\vec{x}) Q(E),
\end{equation}
with $Q(E)$ given by Eq.~\ref{eq:source} and $\rho(\vec{x})$ given
by Eq.~\ref{eq:rho}. This gives us explicitly 
\bea
Q(E,\vec{x}) &=&  1.5 \times 10^{-26} 
\left( \frac{\dot{N}_{100}}{1 \mbox{ century}^{-1}} \right) 
\left( \frac{f_e}{0.15} \right) 
\left(\frac{100 \mbox{ GeV}}{\ecut} \right)^{2-\alpha} \nonumber  \\ 
&& \times \left(\Gamma(0.4)/\Gamma(2-\alpha)\right)
100^{\alpha-1.6}
\left(E/{\rm GeV}\right)^{-\alpha}  e^{-E/E_{cut}}   \\
&& \times e^{-r/(\mbox{4 kpc})-|z|/(\mbox{0.1 kpc}) } \mbox{  GeV}^{-1}
\mbox{ s}^{-1} \mbox{ cm}^{-3}\,, \nonumber
\eea 
for our fiducial $r_0$, $z_0$ and $r_{\rm disk}$ values.

We use GALPROP \cite{Strong:1998pw,Moskalenko:1997gh} to
compute the diffusion, and check the results with an ordinary partial
differential equation solver.  The diffusion coefficient is
\beq
K(E) = K_0 \left( \frac{E}{3 \mbox{ GeV}} \right)^{\delta},
\eeq
where we take $K_0 = 5 \times 10^{28}$ cm$^2$/s and the 
index $\delta = 0.33$ as our fiducial model parameters. We use the
diffusive reacceleration scheme in Galprop with Alfv\'en speed of 36
km/s. These diffusion coefficient values are consistent with those
used in the literature to fit cosmic ray data~\cite{Ptuskin:2005ax}.
The energy loss coefficient $B(E,\vec{x})$
is calculated in GALPROP.  It is dominated by inverse Compton
scattering and synchrotron radiation for electrons and positrons in
the energy range of interest.  The energy loss due to inverse Compton
scattering is calculated using the Klein-Nishina cross section with an
interstellar radiation field that comes with the GALPROP package,
discussed in  \cite{Porter:2005qx,Strong:1998fr}.  

Energy loss due to synchrotron radiation is calculated using an
exponential form of the galactic magnetic field, 
\beq
B(r,z) = B_0 f_C(R)+ (1-f_C(R) )B_\odot
e^{-(r-r_\odot)/r_b-|z|/z_b} \label{eq:B-field}, 
\eeq
where $R^2=r^2+z^2$, $B_0$ is the magnetic field at the
center of the galaxy and $B_\odot=5\mu {\rm G}$ is the local magnetic
field. We choose a characteristic scale $r_b=10$ kpc and $z_b=2$ kpc
to be consistent with earlier studies aiming to explain the Haslam
data and cosmic ray data \cite{Strong:1998fr,Broadbent:1990}. The value
$B_0$ and the function $f_C(R)$ encode our ignorance 
about the magnetic field at the center of the galaxy. Neither the
detailed structure nor the value of the magnetic field in the inner
galactic region is well known~\cite{Han:2006nx}. We choose
$B_0=30\, \mu {\rm G}$ as our fiducial value although smaller and
larger values also lead to consistent fits to the haze spectrum and
its variation with latitude. 
Since there is no observational guidance on the specific structure of
magnetic field within a couple of kpc from the galactic center, we adopt
an exponential form, $f_C(R) = \exp(-R/R_C)$ and set $R_C=1$~kpc. 
We note that this form for $f_C(R)$ achieves the aim of smoothly
transitioning from a spherical magnetic field profile in the inner
galactic region to the cylindrical B field profile that are commonly
used to 
fit Haslam data, cosmic ray data and diffuse galactic gamma-ray
background \cite{Strong:1998fr}. Our adopted magnetic field towards
the center of the galaxy is consistent with the average magnetic field
deduced on scales of order a kpc about the  galactic center from
observations of non-thermal radio filaments~\cite{LaRosa:2005ai}.  

We also note that the 408 MHz Haslam maps do not 
directly constrain the inner galactic magnetic field strengths or
structure since the synchrotron signal depends on $n_e B_\perp^{1.8}$
\cite{Broadbent:1990}, where $n_e$ is the number density of electrons
contributing to the 408 MHz signal and $B_\perp$ is the component of
the magnetic field perpendicular to the line-of-sight. The pulsars do
not contribute significantly to the Haslam map. There must be a different
source of $e^+e^-$ with energies at or below GeV that contribute
dominantly to the Haslam map. Thus, for an assumed magnetic field
spatial distribution, the Haslam maps determine the spatial
distribution of these lower energy electrons.  It is clear from the
Haslam maps that these lower energy electrons must be distributed
more diffusely than the pulsar distribution assumed here. More
detailed work on this subject would require a computation of the
secondary positrons along with primary $e^+e^-$ produced by supernova
remnants, away from the galactic disk. The secondary positrons, which
are sub-dominant at multi-GeV energies in the solar neighborhood, will
contribute significantly at $\sim {\rm GeV}$ and lower energies because
of their steeper spectrum~\cite{Strong:2004de} and there are large
uncertainties in these calculations~\cite{Delahaye:2008ua}.

Our assumptions about energy losses differ from earlier work exploring
constraints with the haze~\cite{Hooper:2008zg}. We do not assume a
spatially constant ratio of energy density of starlight to the
magnetic field energy density to fit the haze data (as discussed also
in Ref.~\cite{Grajek:2008jb}); we parametrize our ignorance  of the
magnetic field in the center of the galaxy directly. This allows us to
make more direct contact with studies aiming to fit cosmic ray and
diffuse photon backgrounds. 

\begin{figure}
  \begin{center}
    \scalebox{0.6}{\includegraphics{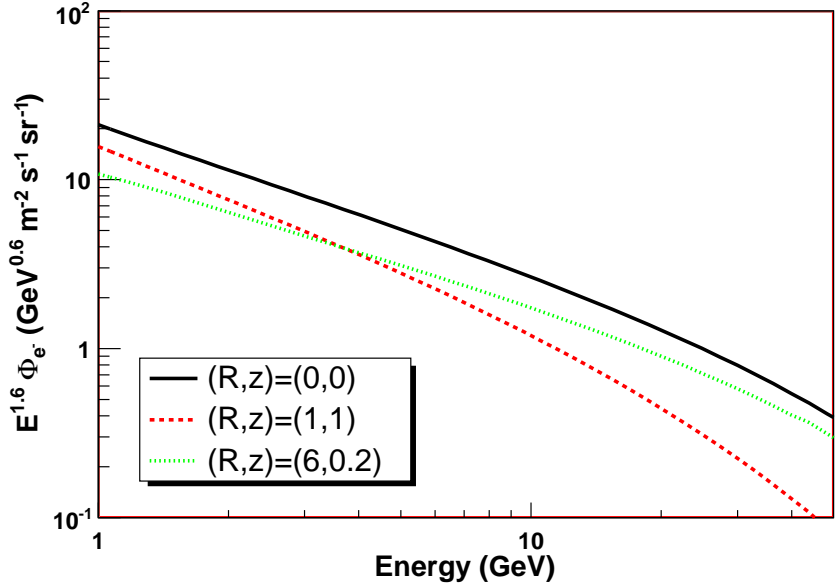}}
  \end{center}
  \caption{The curves show the $e^+e^-$ energy spectrum at different
  locations in 
  the galaxy that contribute significantly to the haze. We see that at
  energies below about 10 GeV, the shape of the spectrum depends on
  the location. For higher energies, the dominant contribution comes
  from more local pulsars. The energies around 10 GeV are particularly
  important for the haze and this figure shows that diffusion and
  energy-loss steepen the energy spectrum index to about -2. 
  We urge
  caution when comparing these estimates to local $e^+e^-$
  measurements as discussed in the text below. \label{fig:Rdep}} 
\end{figure}

We plot in Fig.~\ref{fig:Rdep} the electron flux spectrum at various
positions with respect to the galactic center.  The diffusion softens
the spectrum considerably, implying a larger flux in the lower
frequency bands of the WMAP haze.  The question is then whether the
spectrum remains sufficiently hard to explain the WMAP haze in the all
frequency bands, from 22 to 93 GHz.  We examine this next. 

After propagation,  at any given point in space, the flux in
synchrotron radiation (in erg/s/Hz) in the presence of the magnetic
field is computed according to the relation \cite{Ghisellini:1988} 
\begin{equation}
\epsilon(\nu,\gamma)=\frac{4 \pi \sqrt{3} e^2 \nu_B}{c} x^2 \left(K_{4/3}(x) K_{1/3}(x)- \frac{3}{5} x [K_{4/3}^2(x)-K_{1/3}^2(x)] \right)
\end{equation}
with $x = \nu/(3\gamma^2 \nu_B)$, $\nu_B = eB/(2\pi m_e)$, and $\gamma(E)$ the electron's boost.

The total flux (in kJy/sr) in a given frequency at a given angle from the galactic center is computed by folding the synchrotron power with the electron distribution at any given point.  The flux is then obtained by integrating along the line of sight:
\begin{equation}
\Phi(\nu) = \frac{1}{4\pi}\int_0^{\ell_{\rm max}} d\ell
\int_0^{\infty} dE\, 
\epsilon(\nu,\gamma(E)) \frac{dn}{dE}(r(\ell),z(\ell)), 
\label{flux}
\end{equation}
where $r(\ell) = |r_e-\ell \cos(\theta)|$ and $z=\ell \sin(\theta)$ if
we restrict out attention to angles $\theta$ above and below the
galactic center, the sun is positioned at $r = r_e = 8 {\rm kpc}$
and $z=0$, and $\ell_{\rm max}$ is set by the height of the diffusion
zone. We now turn to presenting our results utilizing the formalism
outlined above.

After propagation, we fit the haze using the flux from Eq.~\ref{flux}
in the all WMAP bands.  The overall normalization of the curve
depends on both the magnetic field, $B_0$ in Eq.~\ref{eq:B-field}, the average pulsar power $W_0$,
and the average pulsar efficiency times pulsar production rate $f_e
\dot{N}_{100}$.  The range for $\dot{N}_{100}$ is 1-3 per century, 
following the rate of core collapse of supernovae in our galaxy. 
\cite{Raffelt:2007nv}.
We have taken $\dot{N}_{100} = 2.8$, following
\cite{FaucherGiguere:2005ny}.  In addition, owing to uncertainties
in the subtraction,  we also allow a constant background at all angles
to be added in the fit.   

The index $\alpha$ in Eq.~\ref{eq:source} and $\ecut$ are important
in fitting the frequency structure of the Haze observed by WMAP.  The
cutoff energy $\ecut$ is required to be above a minimum value
$\gtrsim 40$ GeV such that there will be enough radiation
into the 93 GHz band, however cutoff energies larger than that will
only serve to increase the required average efficiency per pulsar to 
reproduce the haze.  Additionally, a soft spectrum, corresponding to a
large $\alpha$,  will not give rise to a large enough amplitude in the
high frequency bands to reproduce the haze.  
Lastly, the fit in the angular direction results by allowing the 
characteristic distance over which the magnetic field is damped from 
the galactic center, $z_b$ in Eq.~\ref{eq:B-field},  to vary.  

\begin{figure}
  \begin{center}
    \scalebox{0.4}{\includegraphics{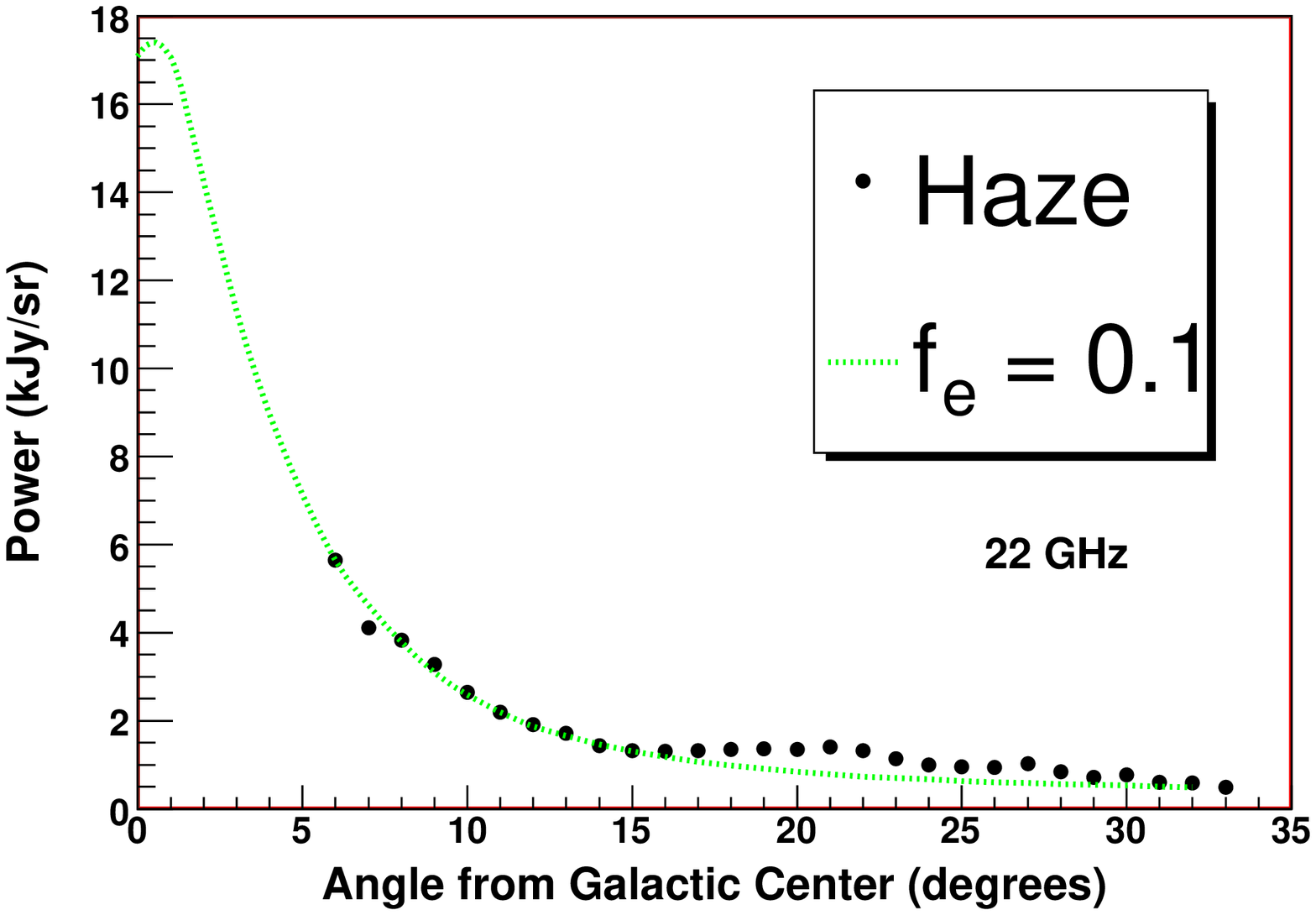}}
    \scalebox{0.4}{\includegraphics{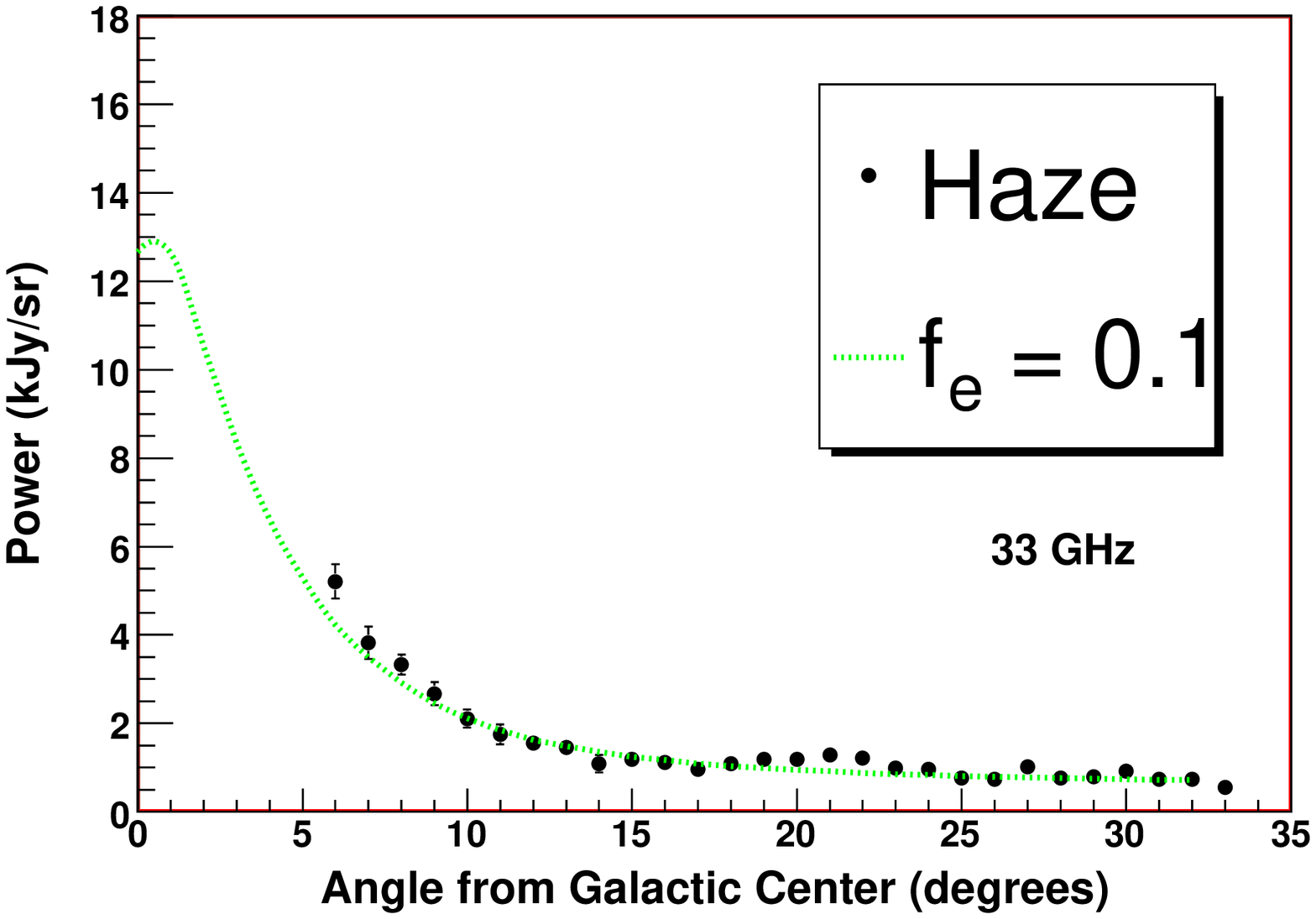}}
    \scalebox{0.4}{\includegraphics{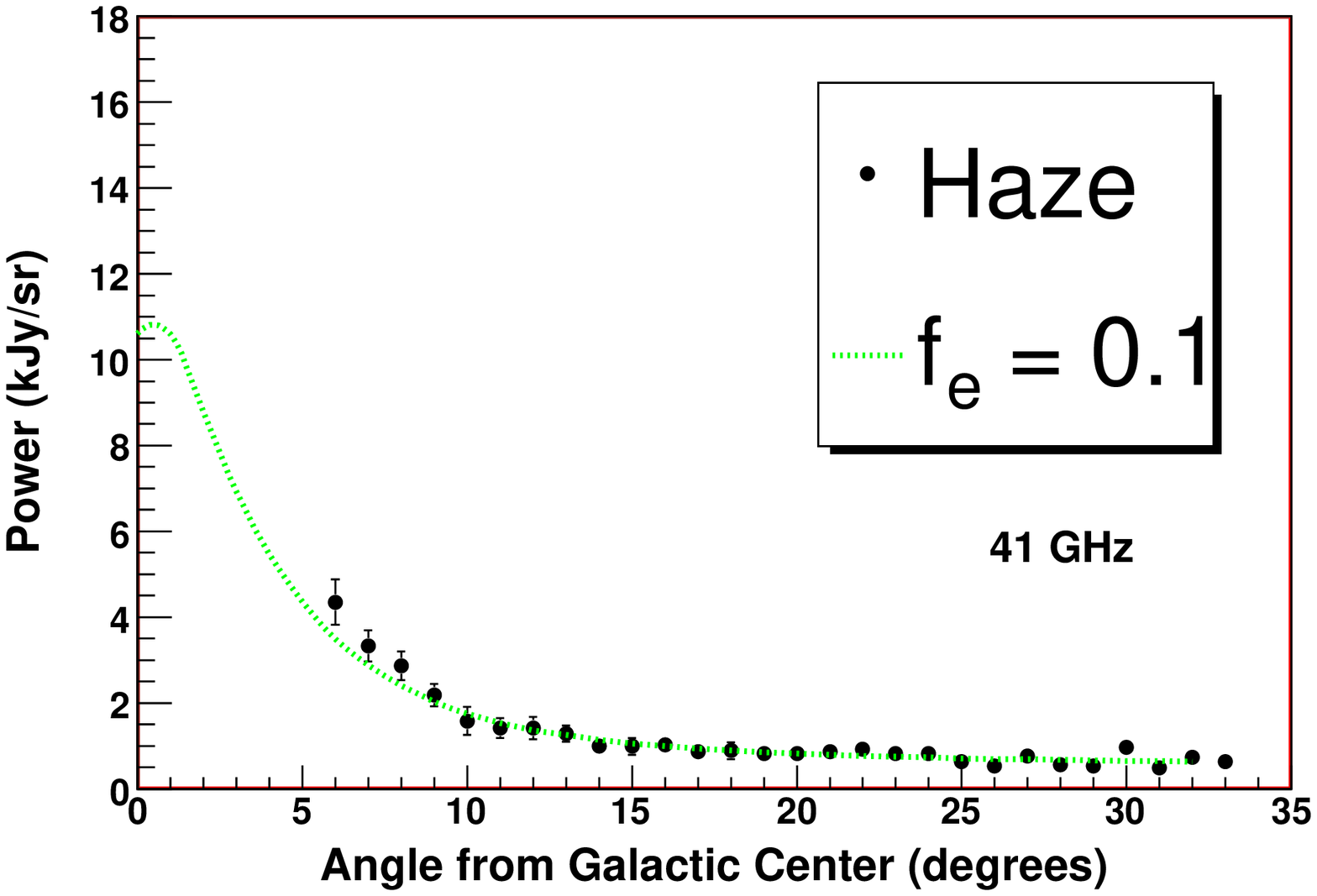}}
    \scalebox{0.4}{\includegraphics{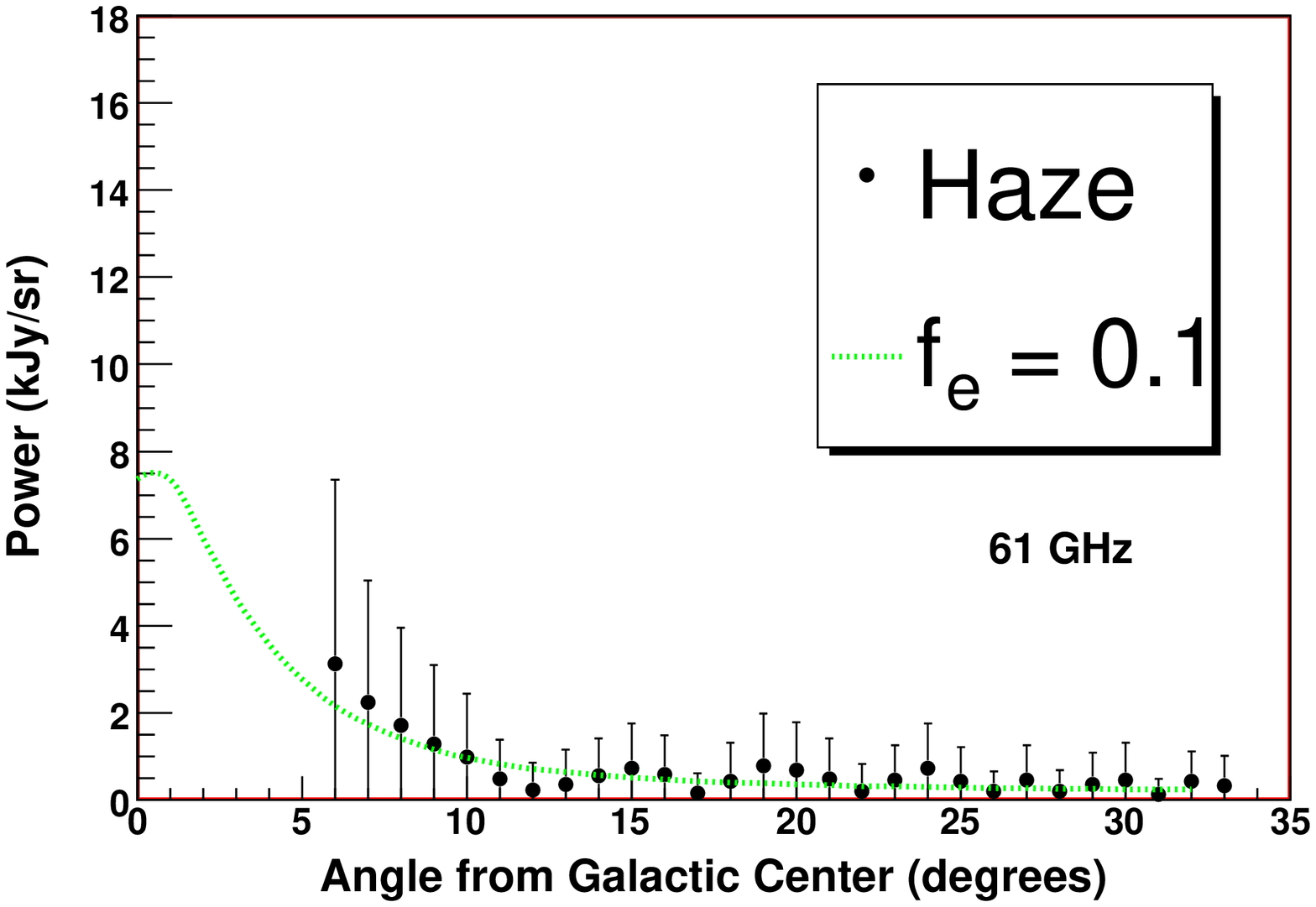}}
    \scalebox{0.4}{\includegraphics{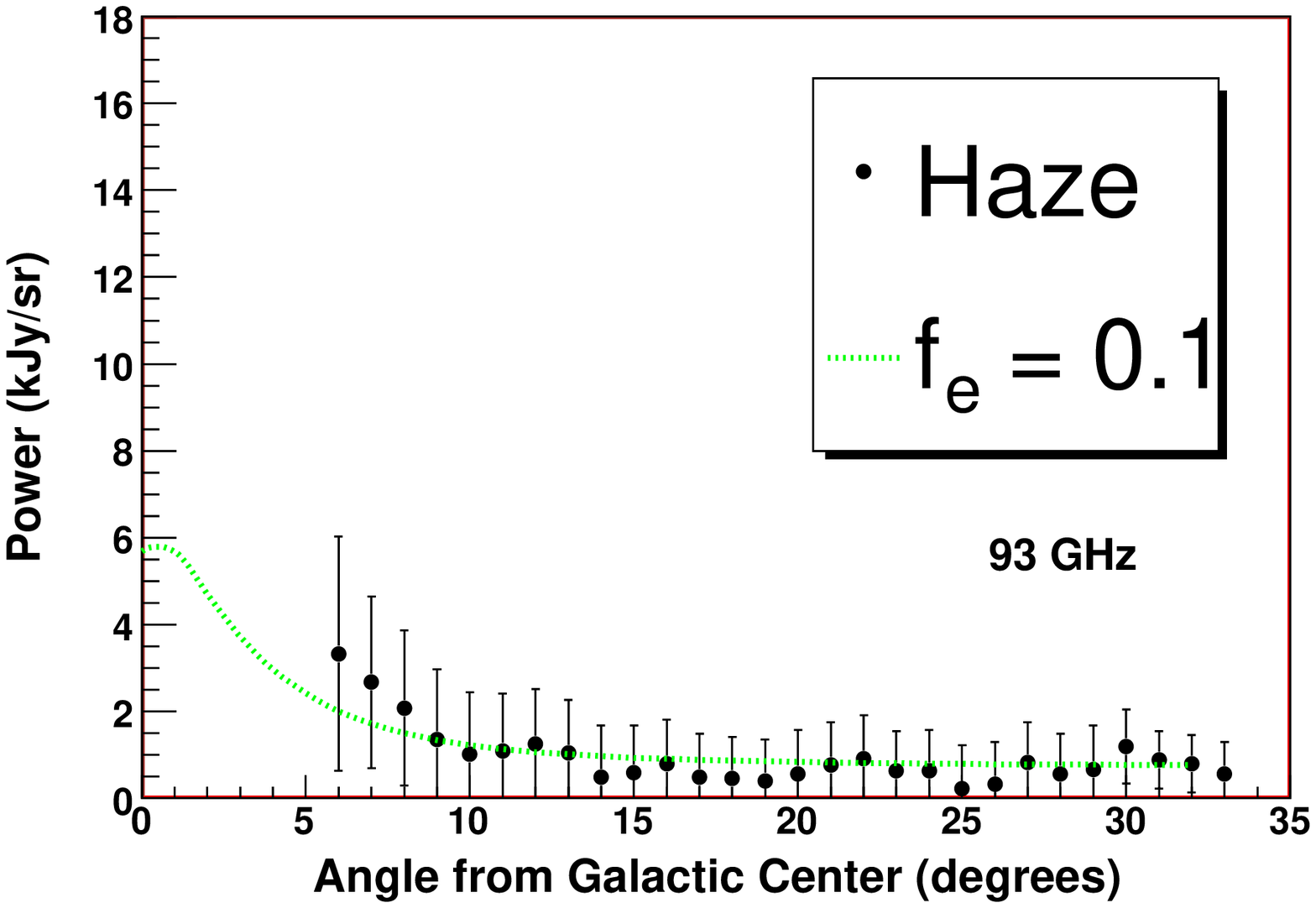}}
  \end{center}
  \caption{WMAP Haze for pulsar injection parameters $\alpha = 1.6$
  and $E_{cut}=100$ GeV and efficiency $f_e=10\%$ (see eqs. (\ref{eq:source}, \ref{eq:Q0def})
  for a definition of efficiency). This efficiency is
  defined as the fraction of the spin-down power converted to $e^+e^-$
  pairs after an assumed maturity age of $10^5 {\rm
  yr}$. The fraction of the total initial pulsar energy required in
  the form of $e^+e^-$ pairs to explain the haze is 3\%. The fits to
  the data above include a floating constant offset for each
  channel and the resulting $\chi^2$ per dof is $1.1$ within the inner
  15$^o$ and $2.1$ within the inner 20$^o$.
  \label{fig:E100alpha1p6}}  
\end{figure}

As shown in Fig.~\ref{fig:E100alpha1p6}, we find that with reasonable
choices of these parameters, a galactic source of pulsars can explain
both the amplitude and frequency dependence of 
the WMAP haze. The plotted synchrotron fluxes are averages over 20 degrees 
in longitude for below the galactic plane.  We find that  $z_b = 2$ kpc gives
a good fit to the angular distribution. In detail, the shape of the
haze depends on both the distribution of the injected pairs as well as
the magnetic field profile. Future Planck data might provide an avenue
to constrain these distributions better.

\begin{figure}
  \begin{center}
    \scalebox{0.4}{\includegraphics{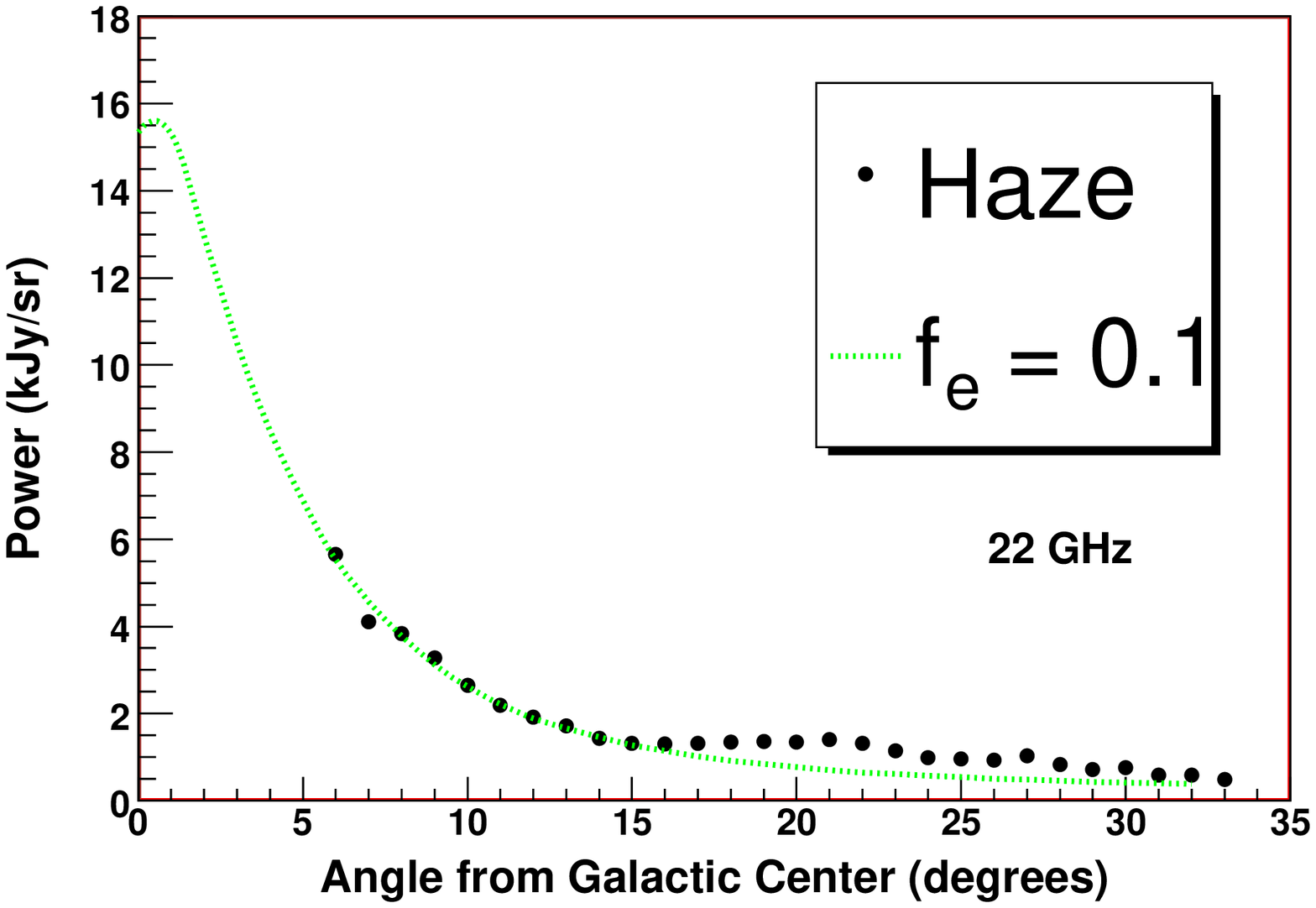}}
    \scalebox{0.4}{\includegraphics{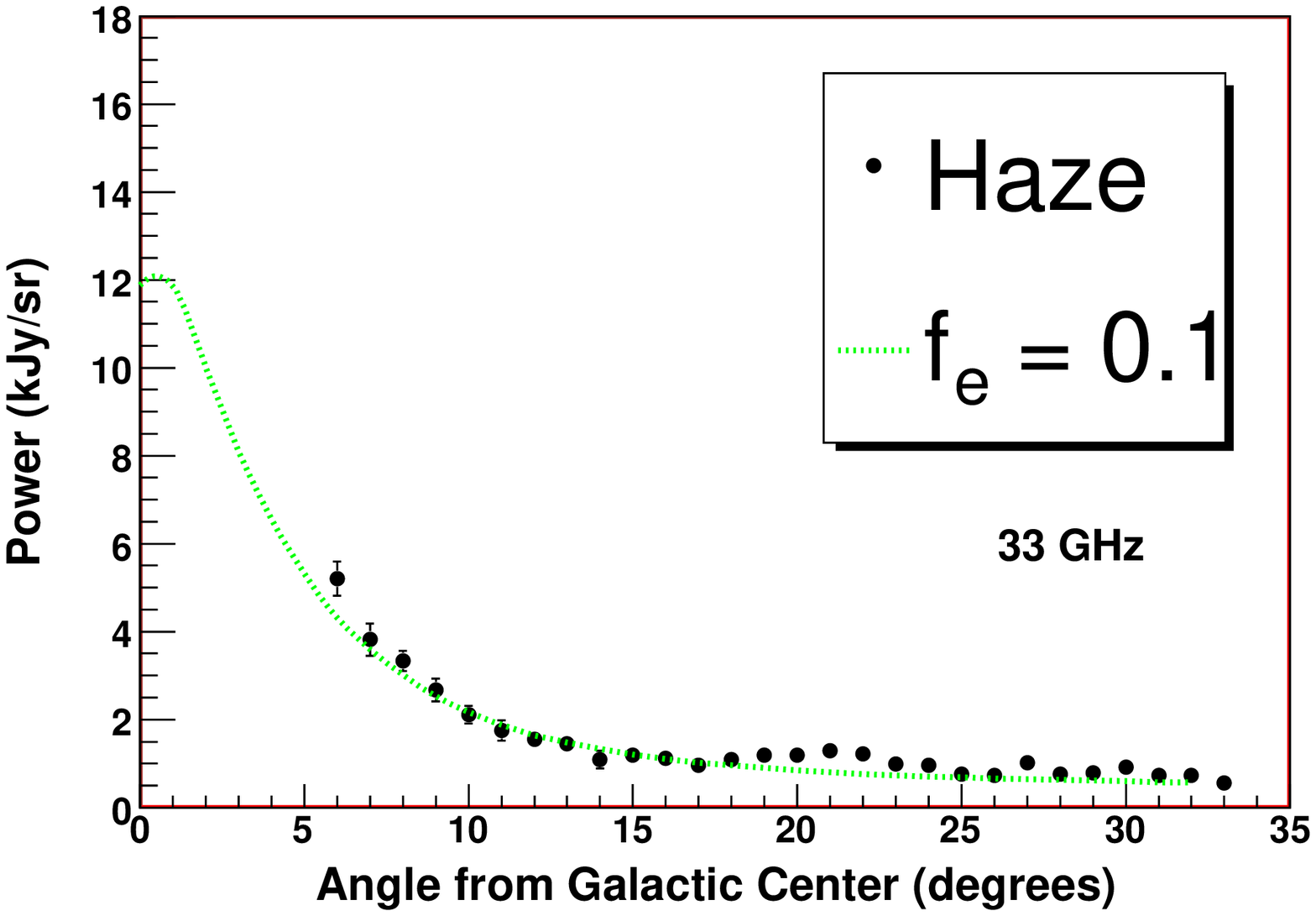}}
    \scalebox{0.4}{\includegraphics{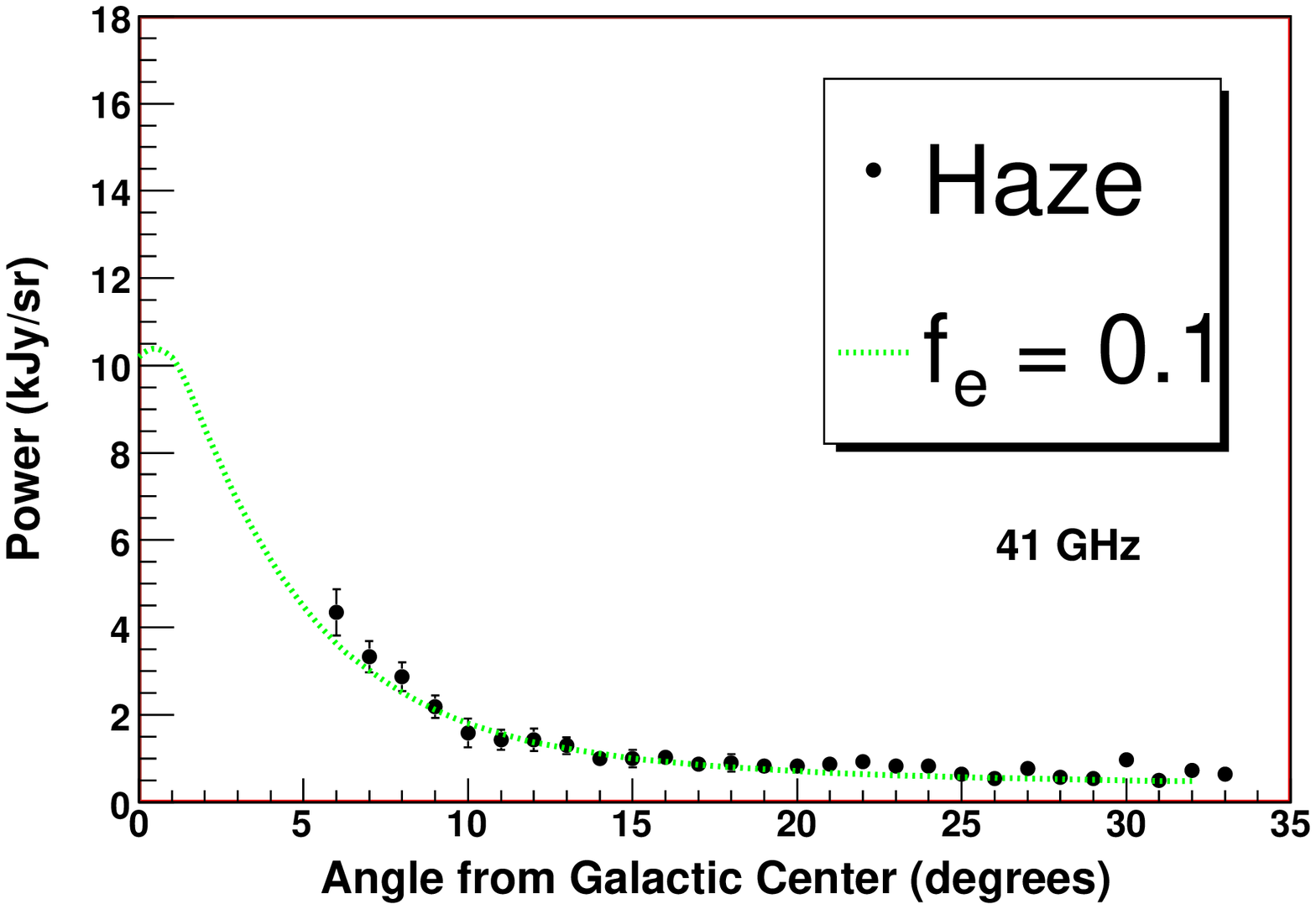}}
    \scalebox{0.4}{\includegraphics{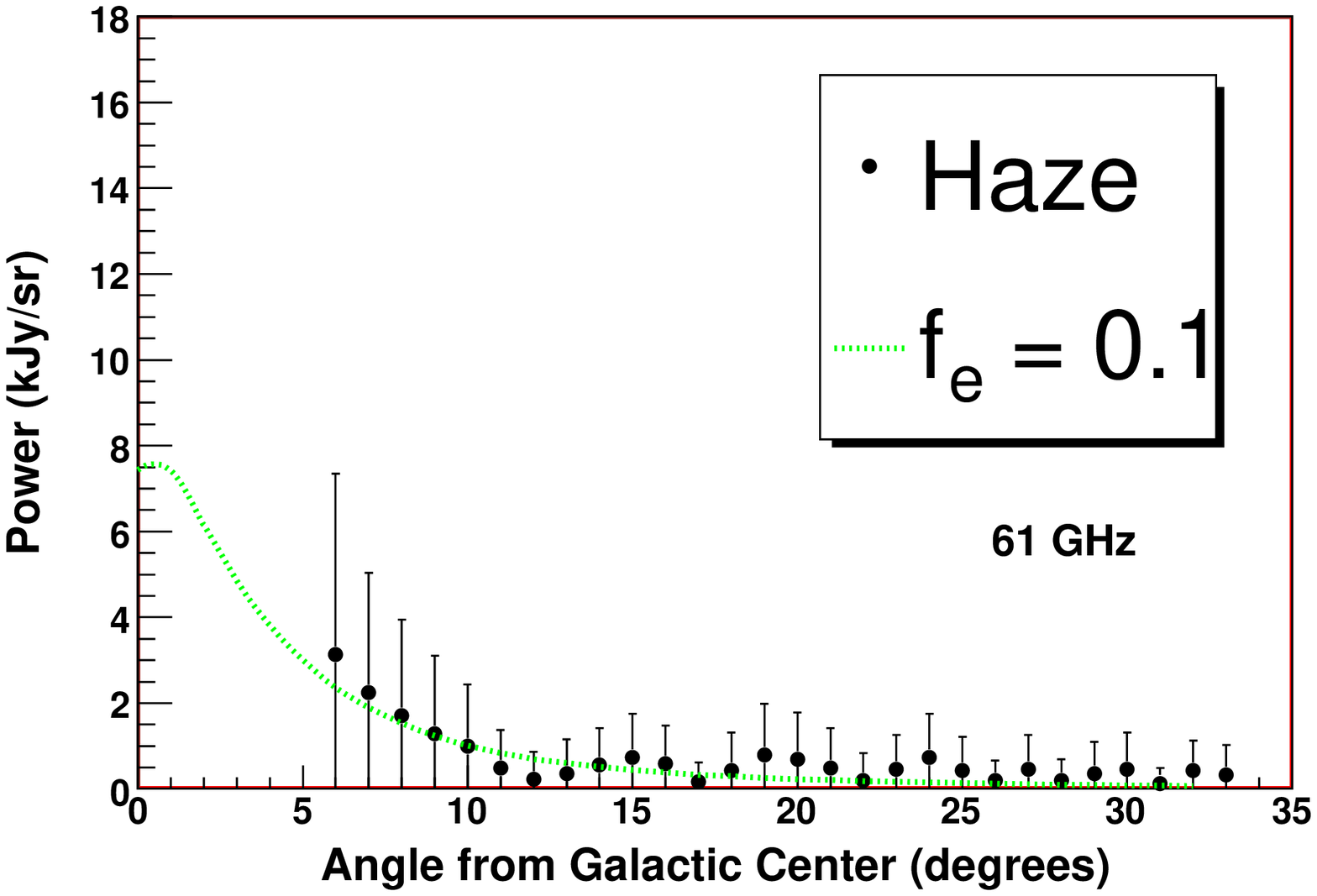}}
    \scalebox{0.4}{\includegraphics{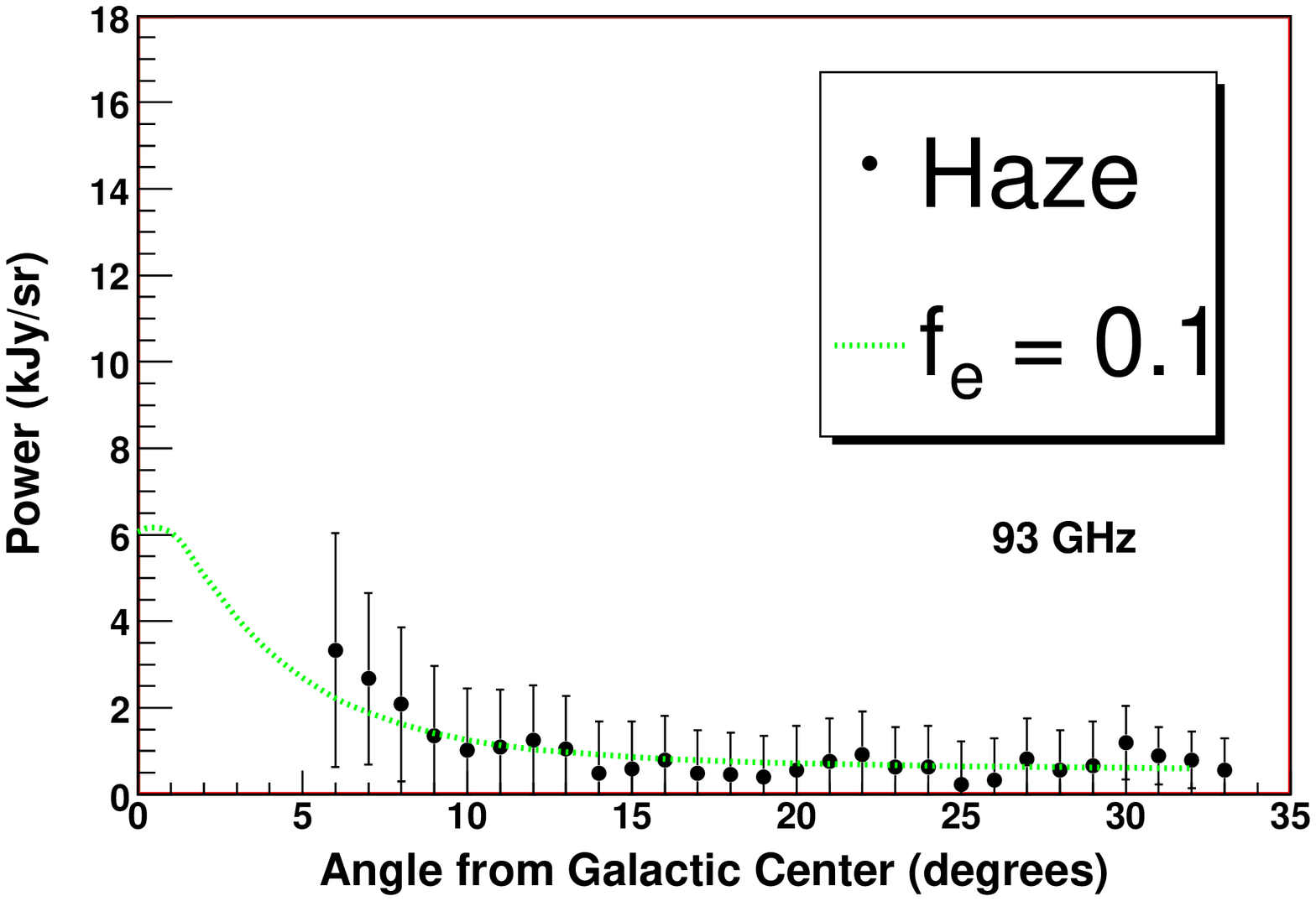}}
  \end{center}
  \caption{WMAP Haze for pulsar injection parameters $\alpha = 1.6$
  and $E_{cut} = 500$ GeV and efficiency $f_e=10\%$ (see eqs. (\ref{eq:source}, \ref{eq:Q0def}) 
  for a definition of efficiency). The larger energy cutoff  results in a
  marginally better fit in the higher frequency bands.  
  The fraction of the total initial pulsar energy required in
  the form of $e^+e^-$ pairs to explain the haze 3\%.   The fits to
  the data above include a floating constant offset for each
  channel and the resulting $\chi^2$ per dof is $0.9$ within the inner
  15$^o$ and $2.3$ within the inner 20$^o$.
  \label{fig:E500alpha1p6}} 
\end{figure}

The cutoff energy $\ecut = 100$ GeV taken from the CZ01
model~\cite{ChengZhang} also  
fits the frequency band requirements of the haze very well.  We show
in Fig.~\ref{fig:E500alpha1p6} the results for a larger choice of
$\ecut$.  Since the parameters $W_0$, $f_e$, and $\dot{N}_{100}$ are interchangeable, 
to get a reasonable efficiency the most likely course is to raise the
average pulsar spin down power. 
In general, the efficiencies
noted there can be substantially and easily lowered by
taking the power per pulsar $W_0$ and the magnetic field at the
galactic center $B_0$ to higher, but still reasonable, values.  In
short, pulsars are a plausible explanation of the haze.  

We also note that the choices of parameters we have made are
consistent with those utilized in \cite{Blasi,Profumo} as a means to
explain the HEAT, AMS, PAMELA, ATIC and PPB-BETs cosmic ray positron
excesses with a single pulsar, or collection of local pulsars,
suggesting that cosmic ray positron and dark matter anomalies may
naturally have the same source.  

We urge caution here that using the local measurements, say
from HEAT and PAMELA, to normalize the pulsar synchrotron contribution
requires many assumptions. 
\begin{itemize}
\item Normalizing the distribution of the pulsars to the local density of
pulsars and then using an analytic form to extrapolate to the center
is not well motivated given the uncertainties in the the deduced
pulsar distribution closer to the galactic center. 
\item For local measurements at energies 5 GeV and higher, the
  dominant contribution to the flux should come from a few local
  sources and we cannot use a smooth source spatial distribution to
  describe this (given the large expected variance). 
\item In the outer gap model for pulsars we use, the energy spectrum
  around a GeV and lower is very uncertain because of the cut-off, as
  discussed earlier. This is not a complicating factor for the haze
  because most of the contribution comes from approximately 5 GeV and higher,
  but it is an issue that must be addressed if one is normalizing to
  lower energy data. In addition, the low energy data is also affected
  by solar modulation. 
\item In order to link the haze to local measurements, we need
  a better understanding of how the $e^+e^-$ flux is affected by
  changes in diffusion between the local and central parts of the galaxy.
\end{itemize}

\subsection{Morphology}

\begin{figure}
  \begin{center}
    \scalebox{0.4}{\includegraphics{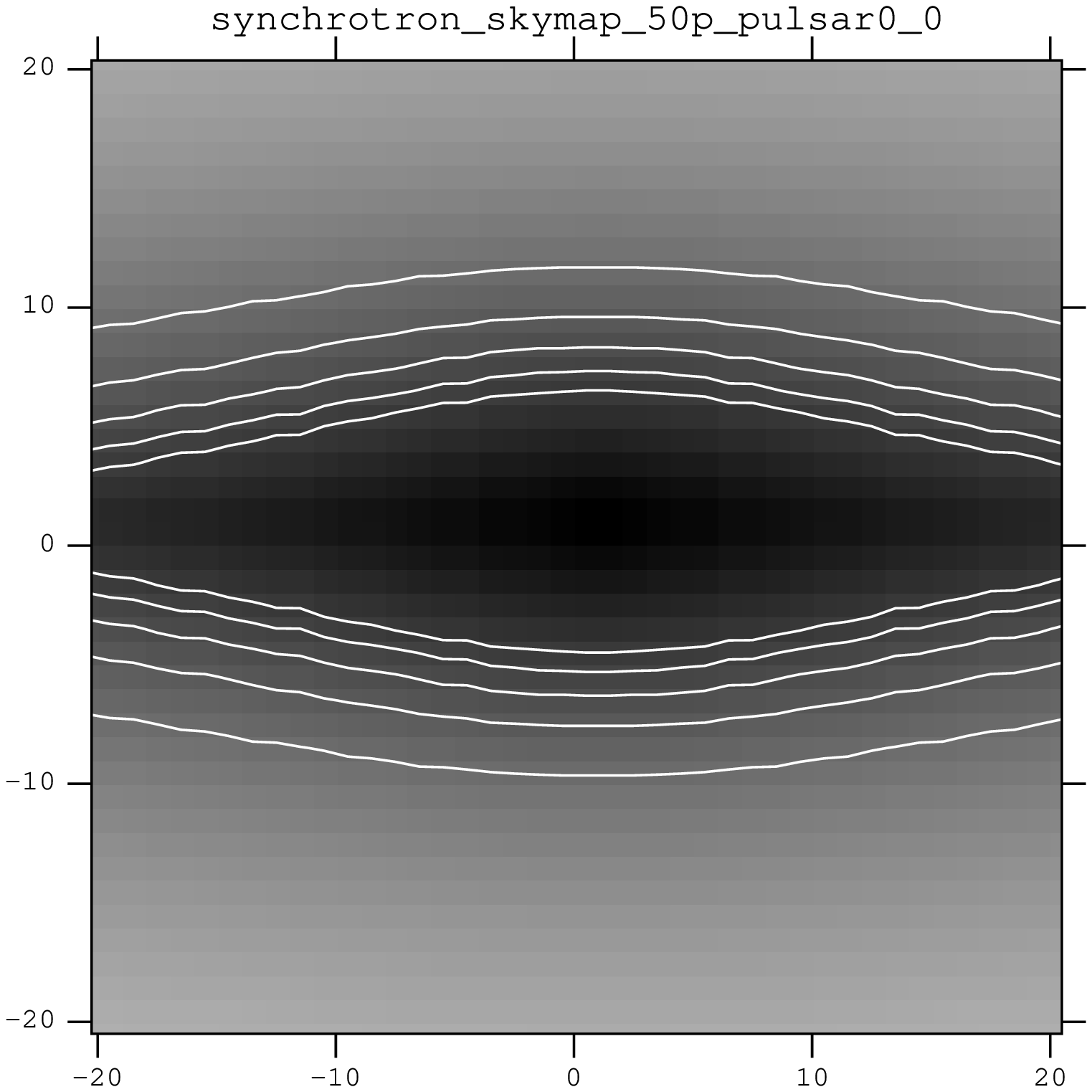}}
    \scalebox{0.4}{\includegraphics{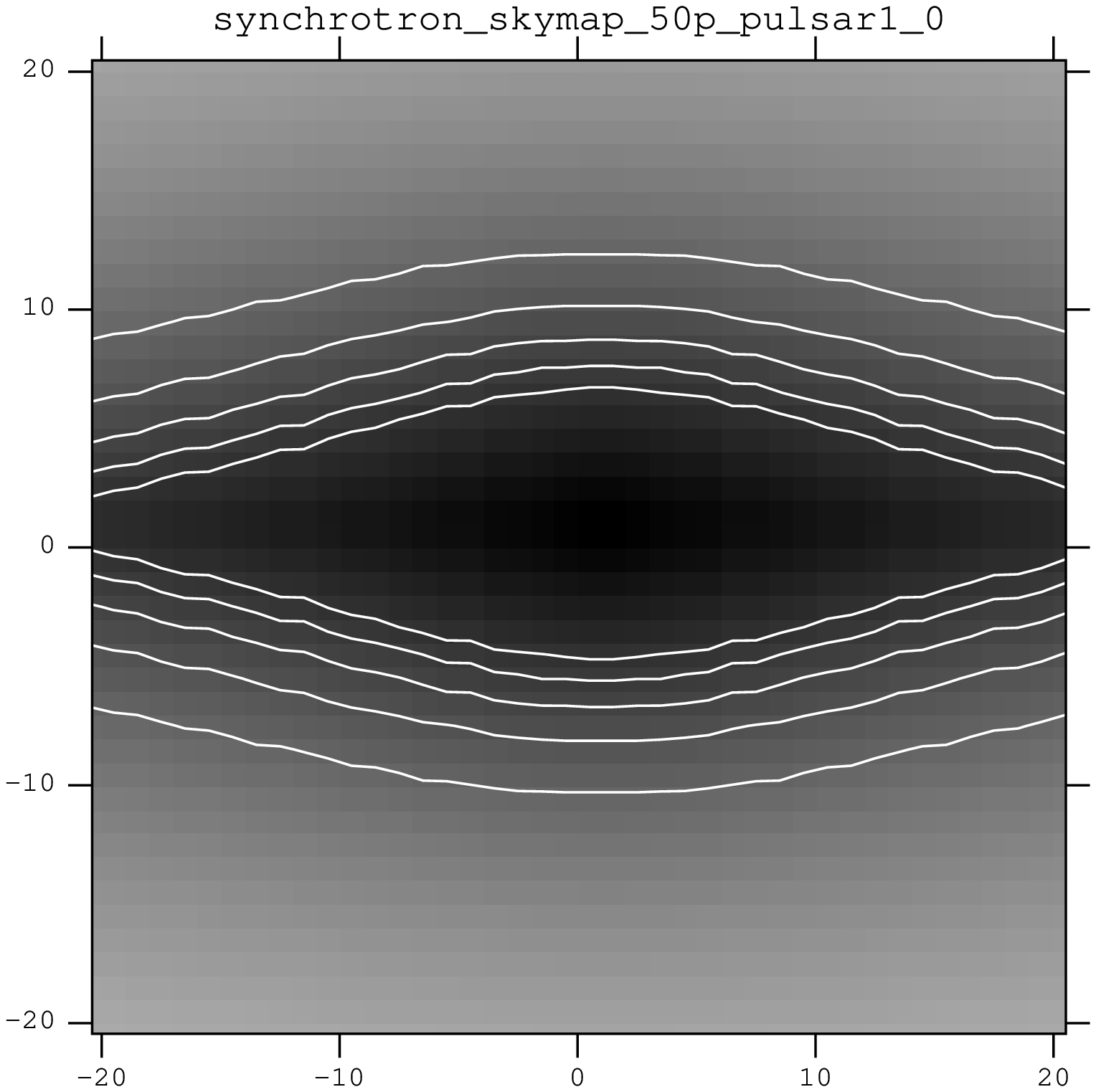}}
    \scalebox{0.4}{\includegraphics{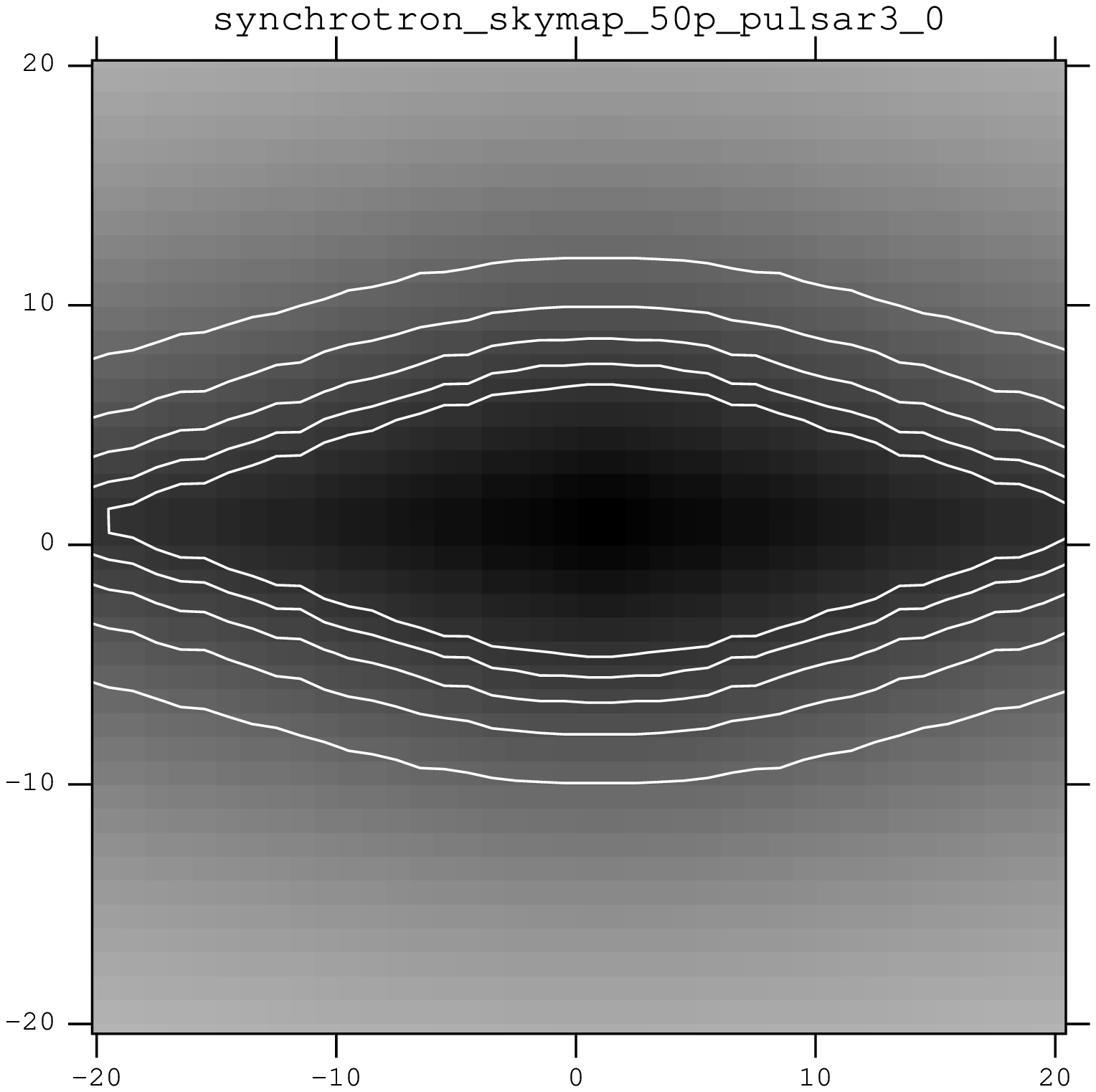}}
    \scalebox{0.4}{\includegraphics{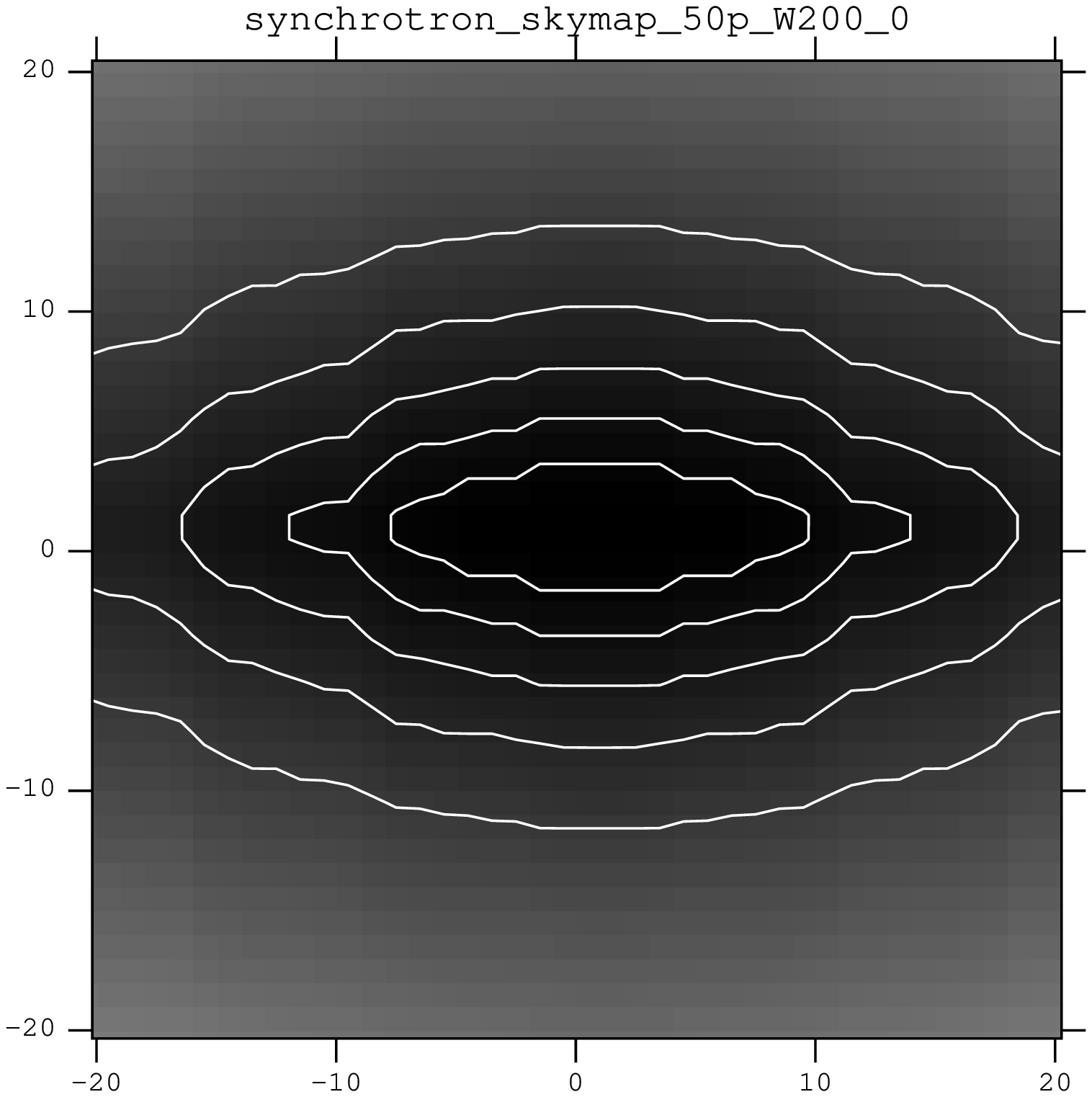}}
  \end{center}
  \caption{22 GHz skymaps in galactic coordinates of the synchrotron
    signal predicted for different models. The contours are spaced
    equally between flux values of 2 and 6 kJy/sr. Note that galactic
    latitudes within roughly 5 degrees of the center are 
    masked while estimating the haze contribution. The left plot in
    the top panel is the model with our fiducial parameters. The plot
    on the right in the top panel utilizes  $K_0 = 1 \times 10^{29}$
    cm$^2$/s (twice the fiducial value) as does the plot on the bottom 
    left. The bottom left plot also modifies the source scale radius
    parameters to  $R_0=3$ kpc.  The plot on the right in the 
    bottom panel is the signal from a model where dark matter is
    annihilating into $e^+e^-$ pairs in the galaxy assumed to have a
    spherically symmetric dark matter density profile $\propto
    r^{-1.2}$ in the inner parts. We have set $f_C(R)=0$ for the dark
    matter plot (see eq. (\ref{eq:B-field})) so  that the effect of the cylindrical galactic
    magnetic field profile  on the signal is clear -- the contours are
    elliptical.  
  \label{fig:skymaps}}  
\end{figure}

The pulsar contribution to the haze is non-spherical, with the
deviation from sphericity determined by a combination of factors
including the diffusion coefficient, the structure of the magnetic
field, and the spatial distribution of the pulsars. Here we vary each
of the these within reasonable range to see their effect on
the morphology of the predicted synchrotron signal. In each case, the
frequency spectrum as well as the detailed drop in average power as a
function of latitude are still good fits to the data. In the top
left panel of Fig.~\ref{fig:skymaps} we show the synchrotron signal at
22 GHz in galactic coordinates  
for our fiducial parameters. The signal is distinctly elliptical, but
judging whether these parameters are disfavored would require an
analysis including the noise properties of the haze map. In the plot to
its right,  we show the signal when the diffusion coefficient is
increased by a factor of 2. This is a modest increase for which we do not
have strong observational constraints on the diffusion coefficient
towards the center of the galaxy. Clearly, the contours look much more
spherical. The largest deviations from sphericity occur close to the
disk where the reconstruction of the haze has the  largest
uncertainties. In the bottom left plot we show the signal when $r_0$
(which governs the radial distribution of the pulsars) is smaller than
the fiducial value and equal to 3 kpc.

As mentioned earlier in this section, the 
population of $e^+e^-$ contributing to the Haslam map is different
from the harder pulsar contribution.  Further, as we have seen, factors 
of two change in the diffusion constant have a significant
effect on  the shape of the synchrotron signal. Thus the behavior of
the diffusion constant in the central parts of the galaxy from energies 1
--10 GeV is important in comparing the Haslam and the haze maps. 
At this level, it is not
clear that diffusion along the disk is the same as perpendicular to
the disk. This will affect the shape of the signal
contours. Lastly, we note that removing the usual (softer) synchrotron
component should also remove some of the more extended features
in the predicted haze maps.   

It is clear that the morphology depends significantly on
parameters that cannot yet be determined from observations. However,
this also opens up the possibility of constraining these parameters
and perhaps the contribution of the pulsars to the haze with Fermi and
Planck data. Note that the ellipticity in the maps trace back both to the
distribution of the pulsars and the cylindrical structure of the
magnetic field. A more detailed comparison to the data would have to
account for the spiral arms and how the galactic magnetic field is
observed to trace the spiral structure.  Note that even a
perfectly spherical source would result in an elliptical haze profile
due to the magnetic field. This is shown in the final plot (bottom
right) of Fig.~\ref{fig:skymaps} where we show the signal resulting
from dark matter annihilations in a halo with inner density profile
scaling as $r^{-1.2}$. For this plot, we have we have set $f_C(R)=0$, so
that the influence of the cylindrical magnetic field structure is
clear -- it makes the contours elliptical.  Conversely, we have shown that the
haze morphology depends sensitively on the assumed diffusion
parameters and the magnetic field in the inner kpc, such that even if
the source of $e^+e^-$ is confined to the disk, the resulting haze can
look approximately spherical.

\section{Conclusions\label{concl}}

We have shown that $e^+e^-$ pairs injected into the interstellar
medium by the galactic pulsars may contribute significantly to the
WMAP haze. The models of pair emission and galactic magnetic fields we 
investigated showed that pulsars could easily account for all of
the haze and successfully reproduce both its angular and frequency
distributions observed in the WMAP data. The parameters for the input $e^+e^-$
spectrum, the magnetic field and inverse Compton energy losses that we
employed to fit the haze towards the galactic center are also
consistent with those that have recently been used to fit the cosmic
ray positron excess from a local distribution of 
pulsars~\cite{Blasi,Profumo}. These facts suggest that both the
haze and the positron excesses may have the same underlying
source. We also considered the morphology of the haze in detail and
pointed out that in it may be possible to use the deviation from
sphericity to bound the contribution of $e^+e^-$ from galactic pulsars
to the haze.

The possibility that the WMAP haze is due to annihilating dark
matter is exciting and it behooves us to search for alternative
astrophysical explanations. More detailed investigations are required
to bracket the contribution of dark matter annihilation products to 
the haze. Correlating various signals of dark matter annihilation
(positrons, anti-protons, anti-deutrons, gamma-rays, haze, etc)
may enable progress in this direction.

\section*{Acknowledgments}

We thank Greg Dobler, Doug Finkbeiner, Jonathan Feng, Dan Hooper, Igor
Moskalenko and Rosie Wyse for discussions and providing helpful suggestions.
This work has been supported by the US DOE grants DE-FG02-95ER40896
(KZ) and DE-FG02-95ER40899 (DP), NASA grant NNX09AD09G (MK) and NSF
grant PHY-0555689 (MK).

\end{document}